\documentclass[lettersize,journal]{IEEEtran}

\usepackage[table]{xcolor}

\usepackage[ruled,vlined]{algorithm2e}
\usepackage{svg}
\usepackage{graphicx}
\usepackage{epstopdf}
\usepackage{amsmath,amsxtra, amsfonts, amssymb}
\usepackage{setspace}
\usepackage{times}
\usepackage[export]{adjustbox}
\usepackage{epsfig}
\usepackage{subcaption}
\usepackage{latexsym}
\usepackage{cite}
\usepackage[hyphens]{url}
\usepackage{tikz,pgf}
\usepackage{pgfplots}
\usepackage{multirow}
\usepackage{breqn}
\usepackage{enumitem}
\usepackage{filecontents}
\usepackage{todonotes}
\usetikzlibrary{arrows}
\usetikzlibrary{automata}
\usetikzlibrary{spy}
\usetikzlibrary{patterns}
\usepackage{tikz-layers}
\usepackage{siunitx}
\usepackage{pgfplotstable}
\usepackage{glossaries-extra}
\setabbreviationstyle[acronym]{long-short}
\GlsXtrEnableEntryCounting
 {acronym}
 {2}

\usetikzlibrary{pgfplots.statistics, pgfplots.colorbrewer} 
\usepgfplotslibrary{statistics}
\usepgfplotslibrary{colorbrewer} 
\pgfplotsset{compat=1.18}

\pgfplotscreateplotcyclelist{MyGroups}{
  {mybox, fill=blue!60},
  {mybox, fill=orange!70},
}

\usepackage{acronym}
\usepackage{comment}
\usepackage{array,tabularx,booktabs}

\usepackage{pifont}
\usetikzlibrary{automata,positioning,arrows.meta,calc}

\usepackage{booktabs,tabularx,pifont}

\newcommand{\base}{\texttt{Baseline}}
\newcommand{\basem}{\texttt{Baseline-\textmu}}

\newcommand{\lean}{\texttt{Lean160}}
\newcommand{\leanm}{\texttt{Lean160-\textmu}}
\newcommand{\leand}{\texttt{Lean160-\textmu+DS}}

\newcommand{\tg}[1]{\texttt{TG\!({#1})}}
\newcommand{\tgm}[1]{\texttt{TG\!({#1})\!-\textmu}}
\newcommand{\tgd}[1]{\texttt{TG\!({#1})\!-\textmu+DS}}

\newcommand{\dsdash}{\texttt{-\textmu+DS}}
\newcommand{\ds}{\texttt{\textmu+DS}}

\definecolor{baseline_short_col}{rgb}{0.29, 0.59, 0.82}
\definecolor{baseline_col}{rgb}{0.68, 0.05, 0.0}
\definecolor{less10_col}{rgb}{0.92900,0.69400,0.12500}
\definecolor{less10_2_col}{rgb}{0.94, 0.88, 0.19}

\definecolor{less30_col}{rgb}{0.59, 0.47, 0.71}
\definecolor{less30_2_col}{rgb}{0.32, 0.09, 0.98}

\definecolor{less60_col}{rgb}{0.55, 0.71, 0.0}
\definecolor{less60_2_col}{rgb}{0.0, 0.5, 0.0}

\newacronym{TV}{TV}{television}
\newacronym{DTTB}{DTTB}{digital television terrestrial broadcasting}
\newacronym{DVB}{DVB}{Digital Video Broadcast}
\newacronym{DVB-H}{DVB-H}{Digital Video Broadcast-Handheld}
\newacronym{ATSC}{ATSC}{Advanced Television System Committee}
\newacronym{ATSC-M/H}{ATSC-M/H}{Advanced Television System Committee - Mobile/Handheld}
\newacronym{IPTV}{IPTV}{Internet Protocol television}
\newacronym{IP}{IP}{Internet Protocol}

\newacronym{UE}{UE}{user equipment}
\newacronym{PC}{PC}{personal computer}

\newacronym{CN}{CN}{core network}
\newacronym{MS}{MS}{mobile station}

\newacronym{ITU-R}{ITU-R}{International Telecommunications Union - Radiocommunication Sector}
\newacronym{IMT-Advanced}{IMT-Advanced}{International Mobile Telecommunications Advanced}
\newacronym{4G}{4G}{fourth-generation of mobile phone communications and Internet access technology}

\newacronym{3gpp}{3GPP}{3rd Generation Partnership Project}
\newacronym{GSM}{GSM}{Global System for Mobile Communications}
\newacronym{UMTS}{UMTS}{Universal Mobile Telecommunications System}
\newacronym{HSPA}{HSPA}{High Speed Packet Access}
\newacronym{lte}{LTE}{Long-Term Evolution}
\newacronym{lte-a}{LTE-A}{Long-Term Evolution Advanced}

\newacronym{e-UTRAN}{e-UTRAN}{evolved Universal Terrestrial Radio Access Network}
\newacronym{eNB}{eNB}{e-UTRAN NodeB}
\newacronym{gNB}{gNB}{gNodeB}
\newacronym{EPC}{EPC}{Evolved Packet Core}

\newacronym{MBMS}{MBMS}{Multimedia and Broadcast Multicast Service}
\newacronym{eMBMS}{eMBMS}{Evolved MBMS}
\newacronym{SFN}{SFN}{single-frequency network}
\newacronym{MBSFN}{MBSFN}{MBMS single-frequency network}
\newacronym{BM-SC}{BM-SC}{Broadcast/Multicast Service Center}
\newacronym{MBMS GW}{MBMS GW}{MBMS Gateway}
\newacronym{MME}{MME}{Mobility Management Entity}
\newacronym{MCE}{MCE}{Multi-cell/multicast Coordinating Entity}
\newacronym{SYNC}{SYNC}{synchronization}
\newacronym{MCCH}{MCCH}{Multicast Control Channel}
\newacronym{MTCH}{MTCH}{Multicast Traffic Channel}
\newacronym{MCH}{MCH}{Multicast Channel}
\newacronym{PMCH}{PMCH}{Physical Multicast Channel}
\newacronym{PDSCH}{PDSCH}{Physical Downlink Shared Channel}

\newacronym{IEEE}{IEEE}{Institute of Electrical and Electronics Engineers}
\newacronym{WiMAX}{WiMAX}{Worldwide Interoperability for Microwave Access}
\newacronym{ASN}{ASN}{access service network}
\newacronym{ASN-GW}{ASN-GW}{ASN gateway}
\newacronym{CSN}{CSN}{Connectivity Service Network}

\newacronym{PA}{PA}{power amplifier}

\newacronym{NI}{NI}{National Instruments}

\newacronym{tdd}{TDD}{time-division duplex}
\newacronym{fdd}{FDD}{frequency-division duplex}
\newacronym{udp}{UDP}{User Datagram Protocol}
\newacronym{APP}{APP}{application}
\newacronym{mac}{MAC}{medium access control}
\newacronym{phy}{PHY}{physical}
\newacronym{RLC}{RLC}{radio link control}
\newacronym{RRC}{RRC}{radio resource control}

\newacronym{FIFO}{FIFO}{first-in first-out}
\newacronym{CRC}{CRC}{cyclic redundancy check}
\newacronym{SAP}{SAP}{service access point}
\newacronym{FEC}{FEC}{forward error correction}
\newacronym{IF}{IF}{intermediate frequency}
\newacronym{RF}{RF}{radio frequency}
\newacronym{mimo}{MIMO}{multiple-input and multiple-output}
\newacronym{aoa}{AoA}{angle-of-arrival}
\newacronym{aod}{AoD}{angle-of-departure}
\newacronym{mcs}{MCS}{modulation and coding scheme}

\newacronym{SPC}{SPC}{superposition coding}
\newacronym{SVC}{SVC}{Scalable Video Coding}
\newacronym{GM}{GM}{generic multicasting}
\newacronym{SCM}{SCM}{superposition coded multicasting}
\newacronym{SIC}{SIC}{successive interference cancellation}

\newacronym{st}{ST}{secondary transmitter}
\newacronym{pt}{PT}{primary transmitter}
\newacronym{ser}{SR}{secondary receiver}
\newacronym{pr}{PR}{primary receiver}
\newacronym{su}{SU}{secondary user}
\newacronym{pu}{PU}{primary user}

\newacronym{awgn}{AWGN}{additive white Gaussian noise}

\newacronym{pdf}{PDF}{probability density function}
\newacronym{ccdf}{CCDF}{complementary CDF}
\newacronym{iid}{IID}{independent and identically distributed}
\newacronym{rf}{RF}{radio frequency}

\newacronym{dd}{DD}{Device-to-Device}
\newacronym{ddu}{DDU}{Device-to-Device user}
\newacronym{dds}{DDS}{Device-to-Device system}
\newacronym{ddt}{DT}{DDU transmitter}
\newacronym{ddr}{DR}{DDU receiver}

\newacronym{bs}{BS}{base station}
\newacronym{bsu}{BSU}{base station associated user}
\newacronym{bss}{BSS}{base station associated system}
\newacronym{bst}{BT}{BSU transmitter}
\newacronym{bsr}{BR}{BSU receiver}

\newacronym{epg}{EPG}{energy per goodbit}
\newacronym{mepg}{MEPG}{modified energy per goodbit}
\newacronym{ee}{EE}{energy efficiency}
\newacronym{se}{SE}{spectral efficiency}

\newacronym{wrt}{w.r.t.}{with respect to}

\newacronym{kkt}{KKT}{Karush-Kuhn-Tucker}
\newacronym{admm}{ADM}{Alternating Directing Method}
\newacronym{cr}{CR}{cognitive radio}
\newacronym{ssi}{SSI}{soft-sensing information}
\newacronym{csi}{CSI}{channel state information}
\newacronym{qsi}{QSI}{queue state information}
\newacronym{el}{EL}{enhancement layer(s)}
\newacronym{snr}{SNR}{signal-to-noise ratio}

\newacronym{NAL}{NAL}{network abstraction layer}
\newacronym{QP}{QP}{quantization parameter}

\newacronym{ofdma}{OFDMA}{orthogonal frequency-division multiple access}
\newacronym{tdma}{TDMA}{time division multiple access}

\newacronym{PUSC}{PUSC}{partial usage of the subchannels}
\newacronym{CFO}{CFO}{carrier frequency offset}
\newacronym{I/Q}{I/Q}{in-phase and quadrature-phase}
\newacronym{ASK}{ASK}{amplitude-shift keying}
\newacronym{PSK}{PSK}{phase-shift keying}
\newacronym{BPSK}{BPSK}{binary phase-shift keying}
\newacronym{QPSK}{QPSK}{quadrature phase-shift keying}
\newacronym{QAM}{QAM}{quadrature amplitude modulation}
\newacronym{PSNR}{PSNR}{peak signal-to-noise ratio}
\newacronym{PELR}{PELR}{packet error and loss rate}
\newacronym{per}{PER}{packet error ratio}
\newacronym{bler}{BLER}{block error rate}

\newacronym{kNN}{\textit{k}-NN}{\textit{k}-nearest neighbor algorithm}
\newacronym{SVM}{SVM}{support vector machines}
\newacronym{nn}{NN}{neural network}
\newacronym{NN}{NN}{neural network}
\newacronym{dnn}{DNN}{deep neural network}
\newacronym{RBF}{RBF}{radial basis function}
\newacronym{RMSE}{RMSE}{root mean squared error}
\newacronym{mse}{MSE}{mean squared error}
\newacronym{lmse}{LMSE}{linear mean square-error estimator}

\newacronym{R2}{$R^2$}{coefficient of determination}

\newacronym{GSA}{GSA}{Global mobile Suppliers Association}

\newacronym{VoD}{VoD}{video on demand}
\newacronym{HEVC}{HEVC}{High Efficiency of Video Coding}
\newacronym{DASH}{DASH}{Dynamic Adaptive Streaming over HTTP}

\newacronym{PUT}{PUT}{people using television}

\newacronym{ADTVS}{ADTVS}{Audience Driven live TV Scheduling}

\newacronym{arq}{ARQ}{automatic repeat request}

\newacronym{harq}{HARQ}{hybrid automatic repeat request}

\newacronym{sdp}{SDP}{semi-definite programming}

\newacronym{tcp}{TCP}{transmission control protocol}

\newacronym{e2e}{E2E}{end-to-end}

\newacronym{ran}{RAN}{radio access network}
\newabbreviation[longplural={radio access technologies}, shortplural={RATs}]{rat}{RAT}{radio access technology}

\newacronym{cran}{CRAN}{cloud radio access network}
\newacronym{udcran}{UD-CRAN}{ultra-dense CRAN}
\newacronym{dran}{DRAN}{distributed radio access network}
\newacronym{hcran}{H-CRAN}{hybrid cloud radio access network}
\newacronym{hetnet}{HetNet}{heterogeneous network}
\newacronym{vcran}{V-CRAN}{virtualized CRAN}
\newacronym{ecran}{E-CRAN}{edge-CRAN}
\newacronym{hvcran}{H-VCRAN}{hybrid-virtualized CRAN}

\newacronym{bbu}{BBU}{baseband processing unit}
\newacronym{rrh}{RRH}{remote radio head}
\newacronym{ru}{RU}{radio unit}
\newacronym{rs}{RS}{remote site}
\newacronym{cs}{CS}{central site}
\newacronym{nr}{NR}{new radio}

\newacronym{rru}{RRU}{radio resource unit}
\newacronym{rb}{RB}{resource block}
\newacronym{hpn}{HPN}{high-power node}
\newacronym{lpn}{LPN}{low-power node}
\newacronym{mabs}{MaBS}{macro basestation}
\newacronym{ue}{UE}{user equipment}

\newacronym{comp}{CoMP}{coordinated multi-point}
\newacronym{ranaas}{RANaaS}{RAN-as-a-Service}

\newacronym{rof}{RoF}{radio over fiber}
\newacronym{wdm}{WDM}{Wavelength Division Multiplexing}
\newacronym{dls}{DLS}{distributed large scale}

\newacronym{qos}{QoS}{quality of service}
\newacronym{qoe}{QoE}{quality of experience}
\newacronym{qee}{QEE}{quality of energy-efficiency}

\newacronym{gg}{GG}{group-to-group}
\newacronym{ht}{HT}{hyper-transceiver}

\newacronym{fh}{FH}{fronthaul}
\newacronym{dl}{DL}{downlink}
\newacronym{ul}{UL}{uplink}

\newacronym{cp}{CP}{Cell-Processing}
\newacronym{up}{UP}{User-Processing}

\newacronym{co}{CO}{center office}

\newacronym{du}{DU}{digital unit}
\newacronym{lc}{LC}{Line-Card}

\newacronym{onu}{ONU}{optical network unit}
\newacronym{olt}{OLT}{optical line terminal}
\newacronym{osw}{OSW}{optical switch}

\newacronym{es}{ES}{ethernet switch}

\newacronym{ppp}{PPP}{Poisson point process}

\newacronym{mppp}{MPPP}{marked Poisson point process}

\newacronym{sinr}{SINR}{signal to noise and interference ratio}
\newacronym{sir}{SIR}{signal to interference ratio}

\newacronym{mbs}{MBS}{macro basestation}
\newacronym{ap}{AP}{access point}
\newacronym{fap}{FAP}{femto-cell access point}
\newacronym{sap}{SAP}{small-cell access point}
\newacronym{iot}{IoT}{Internet of Things}
\newacronym{ti}{TI}{Tactile Internet}
\newacronym{lsm}{LSM}{linear scalarizing method}

\newacronym{lp}{LP}{Low-Priority}
\newacronym{hp}{HP}{High-Priority}
\newacronym{lpu}{LPU}{Low-Priority user}
\newacronym{hpu}{HPU}{High-Priority user}
\newacronym{lps}{LPS}{Low-Priority system}
\newacronym{hps}{HPS}{High-Priority system}

\newacronym{ttm}{TTM}{time to market}
\newacronym{udn}{UDN}{ultra-dense network}

\newacronym{capex}{CAPEX}{capital expenditure}
\newacronym{opex}{OPEX}{operational expenditure}

\newacronym{cpri}{CPRI}{common public radio interface}
\newacronym{otn}{OTN}{optical transport network}
\newacronym{pon}{PON}{passive optical network}
\newacronym{twdm}{TWDM}{time and wavelength division multiplexing}

\newacronym{ec}{EC}{Edge-Cloud}
\newacronym{cc}{CC}{Central-Cloud}

\newacronym{mmw}{m-Wave}{Milli-Meter wave}

\newacronym{gops}{GOPS}{giga operation per second}
\newacronym{mops}{MOPS}{mega operation per second}

\newacronym{ip}{IP}{internet protocol}
\newacronym{rlc}{RLC}{radio link control}

\newacronym{mno}{MNO}{mobile network operator}
\newacronym{mi}{MI}{modulation index}

\newacronym{wifi}{WiFi}{wireless local area network}
\newacronym{cpu}{CPU}{central processing unit}
\newacronym{vcpu}{VCPU}{virtual CPU}
\newacronym{vm}{VM}{virtual machine}

\newacronym{urs}{UrS}{user requested service}

\newacronym{rsf}{RSF}{radio sub-frame}
\newacronym{siso}{SISO}{single-input single-output}

\newacronym{mec}{MEC}{mobile edge computing}
\newacronym{co2}{CO$_{2}$}{carbo dioxide}

\newacronym{ar}{AR}{augmented reality}
\newacronym{vr}{VR}{virtual reality}
\newacronym{cfp}{CFP}{communication function processing}
\newacronym{ptp}{PTP}{precision time protocol}

\newacronym{voip}{VoIP}{voice over Internet protocol}

\newacronym{da}{DA}{data analytics}

\newacronym{kpi}{KPI}{key performance indicator}

\newacronym{fsmc}{FSMC}{finite state markov chain}
\newacronym{ml}{ML}{machine learning}
\newacronym{dgd}{DGD}{distributed gradient descent}

\newacronym{5g}{5G}{fifth generation of mobile communication systems}
\newacronym{gnbcu}{gNB-CU}{gNB central unit}
\newacronym{gnbdu}{gNB-DU}{gNB distributed unit}
\newacronym{ecpri}{eCPRI}{common public radio interface}
\newacronym{fl}{FL}{federated learning}
\newacronym{rsrq}{RSRQ}{reference signal received quality}
\newacronym{rsrp}{RSRP}{reference signal received power}
\newacronym{urllc}{URLLC}{ultra-reliable low latency communications}
\newabbreviation{embb}{eMBB}{enhanced mobile broadband}
\newabbreviation{mmtc}{mMTC}{massive machine type communication}

\newacronym{mae}{MAE}{modified autoencoder}
\newacronym{mtc}{MTC}{machine type communication}
\newacronym{pca}{PCA}{principal component analysis}

\newacronym{cps}{CPS}{cyber-physical system}
\newacronym{gnb}{gNB}{gNodeB}
\newacronym{ref}{REF}{reliability enhancement feature}
\newacronym{nfo}{NFO}{network level feature orchestrator}
\newacronym{dc}{DC}{data center}
\newacronym{nf}{NF}{network function}
\newacronym{vnf}{VNF}{virtual network function}
\newacronym{nfv}{NFV}{network functions virtualization}
\newacronym{nssmf}{NSSMF}{network slice subnet management function}
\newacronym{nsmf}{NSMF}{network slice management function}
\newacronym{ai}{AI}{artificial intelligence}
\newacronym{rl}{RL}{reinforcement learning}
\newacronym{drl}{DRL}{deep reinforcement learning}
\newacronym{ddpg}{DDPG}{deep deterministic policy gradient}
\newacronym{dqn}{DQN}{deep Q-networks}

\newacronym{a2c}{A2C}{advantage actor-critic}

\newacronym{td3}{TD3}{twin delayed deep deterministic policy gradient algorithm}
\newacronym{sgd}{SGD}{stochastic gradient descent}
\newacronym{um}{UM}{unacknowledged mode}
\newacronym{am}{AM}{acknowledged mode}
\newacronym{mdp}{MDP}{Markov decision process}
\newacronym{tti}{TTI}{transmission time interval}
\newacronym{asm}{ASM}{advanced sleep mode}
\newacronym{dss}{DSS}{dynamic spectrum sharing}
\newacronym{FE}{FE}{front end}

\newabbreviation{bsac}{BSAC}{branching \acrlong{sac}}
\newabbreviation{sac}{SAC}{soft actor-critic}
\newabbreviation{inf}{InF}{indoor factory}
\newabbreviation{pmf}{PMF}{probability mass function}
\newabbreviation{cdf}{CDF}{cumulative distribution function}

\newabbreviation{ofdm}{OFDM}{orthogonal frequency-division multiplexing}

\newabbreviation[description={\glsxtrshort{inf}-with dense clutter and high base station height}]{inf-dh}{InF-DH}{indoor factory with dense clutter and high base station height}

\newabbreviation{los}{LOS}{line-of-sight}
\newabbreviation[description={non-\glsxtrshort{los}}]{nlos}{NLOS}{non-LOS}

\newabbreviation{rach}{RACH}{random access channel}
\newabbreviation{prach}{PRACH}{physical random access channel}
\newabbreviation{ssb}{SSB}{synchronization signal block}
\newabbreviation{csirs}{CSI-RS}{channel state information reference signal}
\newabbreviation{sib}{SIB}{system information block}

\newabbreviation{sr}{SR}{scheduling request}
\newabbreviation{bcch}{BCCH}{broadcast control channel}
\newabbreviation{sdu}{SDU}{service data unit}
\newabbreviation{pdcp}{PDCP}{packet data convergence protocol}
\newabbreviation{prb}{PRB}{physical resource block}

\ifCLASSINFOpdf
\else
\fi

\pgfplotsset{compat=1.18}

\begin{document}

\title{Holistic Energy Performance Management:\\ Enablers, Capabilities, and Features}

\author{
\IEEEauthorblockN{Meysam Masoudi\IEEEauthorrefmark{1}, Milad Ganjalizadeh\IEEEauthorrefmark{1},
Tahar Zanouda\IEEEauthorrefmark{2},
P\r al Frenger\IEEEauthorrefmark{1}}\\
\IEEEauthorblockA{\IEEEauthorrefmark{1}Ericsson AB, Sweden.\\
\texttt{\{meysam.masoudi, milad.ganjalizadeh, pal.frenger\}@ericsson.com}
}
\IEEEauthorrefmark{2}\texttt{tzanouda@acm.org}
}

\markboth{Journal of IEEE}%
{ IEEE Journals}
\maketitle

\thispagestyle{plain}
\pagestyle{plain}

\begin{abstract}

Energy consumption is a significant concern for mobile network operators, and to enable further network energy improvements it is also an important target when developing the emerging 6G standard. In this paper we show that, despite the existence of many energy-saving features in 5G \gls{nr} networks, activating them in isolation yields only suboptimal savings and often compromises other network \glspl{kpi} such as coverage or latency. We first introduce a compact taxonomy that distinguishes hardware capabilities from higher-layer features. Features fall into two classes: (i) signaling and scheduling mechanisms that create idle windows, and (ii) features that utilize those windows to save energy. We then present a feature orchestrator as a logical node to coordinate between features to maximize the gain. Using a 3GPP-aligned simulator with product-realistic parameters, we show that coordinating lean \gls{nr}, scheduling, and advanced sleep modes significantly reduces \gls{gnb} energy consumption with negligible throughput loss, compared to the uncoordinated scenario.
We conclude by outlining open issues in observability, system dynamics, coordination, and intelligent automation for energy performance management.
\end{abstract}


\IEEEpeerreviewmaketitle

\vspace{-5pt}
\section{Introduction} \bstctlcite{IEEEexample:BSTcontrol}

The Information and Communication Technology (ICT) sector consumed about $4\%$ of the world’s electricity in its operational phase and contributed roughly $1.4\%$ of global greenhouse gas emissions~\cite{malmodin2024ict,zhang2023towards}. Although 5G is more energy-efficient per bit than 4G, its wider bandwidth, its use of many more antenna branches, and overall, its more capable hardware, have increased absolute energy use ~\cite{standard2022environmental}. 
Therefore, it is crucial to reduce networks' energy consumption and hence emissions and operating costs~\cite{3gppEnergy}.

Mobile networks are dimensioned for the peak hours, leaving substantial capacity underutilized for most of the day. In commercial European 4G deployments, the average cell occupies about $20\%$ of \glspl{prb} \cite{frenger2024energy}; in early 5G \gls{nr} this drops below $6\%$. This under-utilization 
presents an opportunity for savings with limited risk to user experience.
However, radio power consumption does not scale proportionally relative to load. The study in~\cite{frenger2014assessment} reports that traffic increases the energy consumption by up to $7.4\%$. Recent field measurements for operational customer network indicate that this is slightly higher today, at around $10\%$.
Rising traffic erodes peak-hour \gls{ue} throughput, restoring \gls{qos} typically requires more spectrum, sites, or transmit power, which increases energy consumption. 
In lab measurements, a single \gls{ru} may exhibit a significant load dependency, with full-load power usage being roughly double the zero-load value. In live networks, concurrent saturation across sectors is rare, which is why aggregated load dependency rarely exceeds $10\%$. Consequently, energy-saving efforts should target the typical (low-to-moderate) load regime by  {creating longer idle intervals} and  {reducing idle power}. The long-term vision is near-zero power at zero load, with energy scaling proportionally with traffic. 

Energy-saving measures, however, entail trade-offs. Features that reduce energy (e.g., putting some \glspl{gnb} to sleep) can degrade \glspl{kpi} such as coverage, latency, or throughput or raise conflicts \cite{larsen2023toward}, if not coordinated. Therefore, activating a single feature in isolation often yields suboptimal energy savings. 
The studies in~\cite{lopez2022survey, sthankiya2024survey} provide a comprehensive literature review on individual \gls{ran} energy-saving mechanisms (e.g, lean signaling and carrier shutdown), and review \Gls{ml} algorithms for prediction and policy learning. Industry roadmaps in~\cite{alliance2024green, masoudi2019green} outline practical pathways towards less energy-consuming networks, including multi-carrier and multi-band \gls{PA} design, and architecture design. The study in ~\cite{saha2024dynamic} reviews the architectural and spectrum/resource sharing strategies to further widen the idle windows and enable deeper sleep across layers.  
Energy measures are interdependent across the \gls{ran}, core, and data-center domains. Therefore, to amplify the gain, energy-saving features must be orchestrated.
Recent initiatives advocate policy-driven, cloud-native control planes that coordinate cross-layer energy performance. Reference~\cite{ demir2024cell} explores energy-aware orchestration in cell-free massive \gls{mimo}, highlighting the need for feature coordination in virtualized \glspl{ran}.


{\color{black} Unlike prior studies that examine individual energy-saving mechanisms in isolation, this paper leverages coordination to unlock the full energy-saving potential of 5G-and-beyond networks.
Specifically, we (i) introduce a capability–feature taxonomy that distinguishes hardware enablers, idle-window–creating features, and idle-window–utilizing features and their timescales and constraints; (ii) propose a policy-driven feature orchestrator that governs when and how existing features may act, resolves conflicts across layers, and aligns timescales under \gls{kpi} constraints; and (iii) provide a case study, using a 3GPP-aligned simulator, showing that advanced sleep modes achieve better gains when jointly configured and coordinated.}

\section{ from capabilities to Features}\label{sec:features}
In this section, we distinguish between hardware energy-saving capabilities (i.e., what equipment can do) and energy-saving features (i.e., the algorithms/policies that decide when and how to utilize those capabilities).
Capabilities include advanced radio sleep states, processing power control, passive/smart cooling, and multi-band \glspl{PA}. Features fall into two families: (i) signaling/scheduling/lean-\gls{nr} mechanisms that  {create idle windows}, and (ii) \gls{RF}-layer actions (\glspl{asm}, carrier shutdown, massive-\gls{mimo} sleep, dynamic power pooling) that  {utilize those windows} for larger savings. Timescales (Table~\ref{tab:taxonomy}) range from symbol-level $\mu$s, to ms-level scheduling, to tens of ms for deeper sleep, and seconds for orchestration; the main \gls{kpi} risks are coverage, latency, and handover robustness. Table~\ref{tab:taxonomy} summarizes this taxonomy and is reused as a reference.


\subsection{Hardware Energy–Saving Capabilities}\label{sec:hw}
 This subsection explains the principal hardware levers, along with their inherent trade-offs, latency vs energy-saving.

\subsubsection{Radio Sleep Capabilities}\label{sec:radio-sleep}
Modern 5G radios offer three built-in sleep levels: micro, light, and deep. Each deeper level saves more power but takes longer to wake-up from.  {\color{black} Higher-level algorithms choose which level to enter based on the length of the idle intervals.}

\begin{itemize}
\item \textbf{Micro sleep}: disables fast-recovery blocks, i.e., \glspl{PA}, low-noise amplifiers, for {\color{black} tens of microseconds}, e.g., approximately 71 \(\mu\text{s}\), between scheduled symbols.
\item \textbf{Light sleep}: deactivates additional \gls{RF}/baseband components for \(1\text{–}5\)\, ms during brief idle gaps, saving energy at the cost of millisecond wake-up delay.
\item \textbf{Deep sleep}: delivers the highest savings. It deactivates most of the \gls{RF} and baseband components, retaining only timing keep-alives; wake-up takes \(10\text{–}50\)\, ms or more. 
\end{itemize}


\subsubsection{Multi-band PA Capability}\label{sec:mb-pa}
\Glspl{PA} are among the most energy-hungry blocks in a \gls{ru} \gls{FE}. A multi-band \gls{PA} employs reconfigurable \gls{RF} networks to operate across multiple bands used by 4G and 5G, consolidating functionality into fewer chains. This capability enables power pooling, where it dynamically reallocates power between bands. 
Transitions occur on millisecond timescales and must be coordinated with the scheduler and traffic steering to meet per-band linearity, filtering, and emission constraints.


\subsubsection{Processing Power Management}\label{sec:proc-power}
Baseband and edge-cloud processors pair {dynamic voltage and frequency scaling} (DVFS) with {dynamic resource sleeping} (DRS).  DVFS lowers clock and core voltage under light load, while DRS places unused cores or accelerators into progressively deeper power-saving modes.  Combined policies add only micro to millisecond latency. Furthermore, function virtualization may consolidate workloads onto fewer active servers, allowing the remainder servers to enter deep sleep and save energy.

\subsubsection{Passive and Smart Cooling}\label{sec:cooling}
Active thermal management infrastructure represents approximately \(30\%\) of a site’s energy consumption \cite{masoudi2019green}.  Two-phase evaporative passive cooling systems, and optimized airflow channels enable passive removal of heat without compressors or fans, driving cooling power toward zero.  When active cooling system remains, AI-based controllers throttle pumps and fans in proportion to load and ambient temperature, which may lead to energy savings.

\subsection{Creating Idle Time-Frequency Windows}\label{sec:createWin}
This subsection reviews the methods that  {shape traffic and signaling}, in time and frequency, to enable longer and wider contiguous idle intervals. We cover three feature families: (i)  {lean \gls{nr} operation} that reduces always-on transmissions; (ii)  {control-/data-plane scheduling} (paging alignment, batching, traffic gating) that time-shifts non-urgent traffic; and (iii)  {traffic steering/load placement} that moves \glspl{ue} to more efficient layers or cells, increasing idle time elsewhere. In isolation, without supporting capabilities in the underlying hardware, these features can yield modest savings. However, they enable opportunities for deeper sleep states to be used.


\subsubsection{Lean \gls{nr} Design}
The 5G \gls{nr} standard enhances energy efficiency in radio networks by incorporating support for sleep states in network equipment. It achieves this through significantly longer periods between mandatory transmissions (e.g., up to $160$\,ms for \gls{prach} period, compared to 4G's $20$\,ms). Additionally, 5G \gls{nr} reduces the need for constant signaling transmissions in the frequency domain, reflecting its ultra-lean physical layer design. This design enables deeper and more extended sleep periods during periods of low or no traffic, resulting in lower energy consumption compared to 4G cells for the same traffic volume.

\subsubsection{Control‑ and Data‑Plane Scheduling}\label{sec:cpdp-sched}

A mobile network operates over two logical planes. The  {control plane} exchanges signaling (e.g., paging, \gls{RRC} messages, mobility commands)  between \glspl{ue}, \glspl{gnb}, and the core; the data plane carries \gls{ue} payloads. Under low traffic, control signaling can dominate energy consumption, especially when frequent paging bursts keep nodes awake.
As the number of \glspl{ue} grows, the cumulative signaling may prevent radios and processors from entering sleep states. Energy‑aware scheduling mitigates this effect by batching control messages, aligning, and extending control signaling periods. 
While the control plane shapes the signaling time, the data plane determines how \gls{ue} traffic is mapped to the time and frequency resources of the air interface. Energy-aware schedulers accumulate best-effort packets into short, high-throughput bursts, then leave the carrier idle long enough to trigger micro-, light-, or even deep-sleep in the radio. On the \gls{ul}, configured-grant and buffer-status-report timers are staggered so \glspl{ue} request resources in aligned slots, further extending contiguous idle periods. Delay-tolerant flows (e.g., video streaming service) can be gated for tens of milliseconds, whereas latency-critical bearers bypass gating to meet 5G \gls{qos} budgets. By matching burst duration and gate timers to each \gls{qos} class, the scheduler both preserves user experience and creates the long, predictable idle windows needed for deeper sleep states (see Section \ref{sec:consumeWin}).


\subsubsection{Traffic Steering}\label{sec:trafficsteer}
Traffic steering dynamically redistributes \gls{ue} traffic among neighboring cells, carriers, and \glspl{rat}. By shifting flows to cells operating on wider, energy-efficient 5G carriers (often $100 \mathrm{MHz}$ in mid band) or to less loaded layers, operators can both increase cell idle time and lower joules per bit, in some deployments by up to an order of magnitude. The method also exploits the fact that more than $80\%$ of \gls{dl} traffic originates indoors: relocating indoor \gls{ue} from outdoor macro layers to indoor small cells satisfies throughput and latency targets at a fraction of the transmit power.
Besides frequency layer migration, modern steering frameworks account for multi-cell sets, diverse \gls{qos} classes, and evolving interference patterns. When coupled with carrier aggregation, steering can balance load while keeping additional carriers in deep sleep, further extending energy gains. Steering decisions must be orchestrated with other energy-saving functions (e.g., \glspl{asm}) to avoid \gls{kpi} degradations.

\subsubsection{Architecture Split}
In 5G, \glspl{gnb} can be divided into a Central Unit (CU) and a Distributed Unit (DU). Centralized processing provides extensive benefits such as joint processing of radio signals, load balancing, and energy saving. The split architecture distributes the processing tasks between these two units. CU handles higher-layer protocols, such as non-real-time processing, including the \gls{pdcp} and \gls{RRC}. DU manages real-time lower-layer functions, such as the \gls{RLC}, \gls{mac}, and parts of the \gls{phy} layer. By splitting the \gls{gNB} into CU and DU, the network can achieve the following energy-saving benefits:
\begin{enumerate}
    \item Flexible Resource Allocation: Splitting allows for a flexible allocation of processing resources, reducing the need for over-provisioning and thus saving energy.
    \item Load Balancing: Traffic can be dynamically balanced between CUs and DUs, ensuring efficient resource use and reducing unnecessary power consumption.
\end{enumerate}



\subsubsection{\Gls{dss}}
To achieve cost-effective 5G coverage, operators also need to use low bands that are currently occupied by 4G. Full spectrum reallocation is often impractical. \gls{dss} allows 5G to utilize existing 4G carriers without disrupting live services. \gls{dss} lets 5G and 4G share the same carrier. Rather than statically reallocating spectrum, the scheduler assigns \glspl{prb} to 5G or 4G every \gls{tti} to track the \glspl{ue} and traffic mix. Coexistence relies on 5G rate-matching around 4G reference and control signals, and careful placement of 5G synchronization so it does not collide with 4G broadcast/multicast subframes, resulting in per-\gls{tti} \gls{dl}/\gls{ul} \gls{prb} partitioning without re-planning carriers.
From the energy perspective, \gls{dss} can reduce network power by avoiding new sites/carriers and consolidating traffic on existing bands. However, 4G’s always-on signals fragment idle periods, and thus, may limit activation of deeper SMs. 

\subsection{Utilizing Idle Time-Frequency Windows}\label{sec:consumeWin}
This subsection explains features that utilize the idle intervals created in Sec.~\ref{sec:createWin}. The design goal is to maximize energy savings per unit idle time through  coordination with other features, e.g., scheduling. 
\subsubsection{Advanced Sleep Modes}\label{sec:asm-feature}
Modern 5G radios expose three graded sleep states in 3GPP known as \glspl{asm} \cite{3GPPTraffic}. As  {capabilities} (Sec.~\ref{sec:radio-sleep}), they define what hardware can switch off and the corresponding wake-up latencies. As a {feature}, the controller decides  {when/how} to enter/exit each level based on traffic and coordination signals. \Glspl{asm} rely on the scheduler to {both create} and {protect} the contiguous idle intervals needed for deeper sleep, and to restore service after wake-up. Moreover, predictive wake-ups  may reduce \gls{kpi} degradation risk. 

\subsubsection{Carrier Shutdown}\label{sec:carriershutdown}

Carrier shutdown (CS) deactivates capacity carriers (i.e., additional cells deployed on top of the always‑on coverage layer) during periods of light traffic. Once the real‑time load on a capacity carrier falls below a configurable threshold for a certain time, the carrier first enters \textit{cell‑sleep}, scheduling is halted and the \gls{PA} bias is lowered while reference signals remain, enabling sub‑$10$\, ms wake‑up. If low load persists, the radio transitions to \textit{deep‑sleep}, powering down most \glspl{PA}, \gls{RF} chains, and parts of the baseband, cutting per‑carrier power by up to \(90\%\). Three parameters dominate the CS design: (i) the load threshold, (ii) minimum on/off duration, and (iii) predictive wake‑up triggers based on traffic forecasts and coverage overlap. Machine learning predictors can reduce  {false sleeps}, premature shutdowns followed by rapid re‑activation. However, coordination is essential with traffic steering (Section~\ref{sec:trafficsteer}). Without such harmonization, aggressive shutdowns risk handover bursts, throughput loss, and hardware wear from excessive power‑cycling.


\subsubsection{Massive \gls{mimo} Sleep}
Massive \gls{mimo} radio uses a large number of antennas up to $64$ transmit ports, each with its own \gls{PA}, \gls{RF} chain, and signal processing. Massive \gls{mimo} sleep disables surplus branches when spatial‑multiplexing demand is low. A typical policy scales a \(64\) array down to \(32\) or \(16\) once the cell load stays below a certain threshold for a certain time, resulting in reduced power consumption. With massive \gls{mimo} sleep activated, \glspl{ue} may start to report less favorable channel state information (CSI), and the scheduler may reduce modulation or allocate more bandwidth. However, if the scheduler is aware of the massive \gls{mimo} sleep, it can properly allocate radio resources and form beams with the available antennas.
While Massive \gls{mimo} Sleep can dramatically reduce radio power consumption, it comes with trade-offs. Deactivating antenna elements may lead to some \gls{kpi} degradations, especially for cell-edge \glspl{ue}. Machine learning predictions of traffic patterns can be used to optimize this process, allowing the system to proactively activate sleep mode without causing transient service impacts ~\cite{wesemann2023energy}.

\subsubsection{Dynamic Power Pooling}\label{sec:dyn-pool}
Conventional radios use a dedicated, peak-dimensioned \gls{PA} per carrier/band, leaving many amplifiers underutilized at light load. {Dynamic power pooling} replaces this with a shared \gls{PA} budget across bands using the multi-band \gls{PA} capability (Sec.~\ref{sec:mb-pa}). Because band traffic peaks seldom coincide, power can be shifted to the busy band on demand; if several bands peak simultaneously, the controller either activates additional \glspl{PA} or accepts minor per-band \gls{kpi} degradation to stay within the budget. Reports indicate that migrating from single-band to pooled multi-band radios (two–three bands per unit) can yield on the order of \(20\text{–}30\%\) site-energy savings. Pooling operates on ms timescales and should be coordinated with scheduling/steering, \gls{asm} hysteresis, and carrier-shutdown sequencing to avoid on/off toggling and maximize contiguous sleep.



{\color{black} Hardware and feature constraints limit achievable energy savings. Individual features expose only local capabilities and lack global network visibility and coordination. A feature orchestrator aggregates this information to enable a coordinated, network-wide decision-making, as described in Section~\ref{sec:orchestrator}.}

\section{Feature Orchestration}\label{sec:orchestrator}
{\color{black}
Mobile networks are designed to provide seamless service and meet stringent \gls{kpi} requirements (e.g., coverage, latency, throughput), while minimizing operational costs, including energy consumption. Energy-saving features often optimize local objectives and, when activated independently, can conflict or overlook cross-layer synergies, leading to inefficiencies or \gls{kpi} degradation. A policy-driven feature orchestration layer is therefore required to coordinate features and translate target \glspl{kpi} into coherent end-to-end behavior.
The feature orchestrator is a logical coordination node that, within the 3GPP architecture, can be implemented in the management and orchestration layer and, within the O-RAN architecture, maps to the service management and orchestration framework. The feature orchestrator ingests (i) operator policies (e.g., energy minimization subject to \gls{kpi} constraints), (ii) network state, and (iii) the set of available features and hardware capabilities. Based on these inputs, it determines which features may act and when, while leaving the actual actuation to the respective feature functions. The orchestrator can be realized using rule-based policies and predictors, or via \gls{ml}-based optimization, which may improve confidence margins.
The orchestrator operates across three functional phases: preparation, execution, and monitoring.
\subsubsection{Preparation Phase}  
The orchestrator aggregates information including operator policies, required \gls{kpi} thresholds, available features and capabilities, and the current network context. It runs a load estimator and short-horizon traffic predictor to estimate expected idle windows, traffic burstiness, and potential load shifts due to mobility. Each feature exposes metadata to the orchestrator, such as its operational timescale, wake-up latency, required inputs, energy-saving potential, and known dependencies or conflicts with other features. Using this information, the orchestrator determines a subset of features that can be safely coordinated under the prevailing conditions. Conflicting combinations (e.g., carrier shutdown without prior traffic steering) are avoided by design.
This selection can be implemented using a feature dependency and conflict graph. The orchestrator also aligns feature timescales by configuring thresholds, dwell timers, and hysteresis parameters, ensuring features are aligned. At the end of this phase, the orchestrator produces a coordinated configuration plan for the selected features.
\subsubsection{Execution Phase}  
The orchestrator issues configuration and activation signals to the selected features through appropriate management and control interfaces. These signals may include enabling or disabling features, setting thresholds or dwell times, and defining priorities or rank constraints among features.
The orchestrator does not directly trigger physical actions such as sleep transitions. Instead, it constrains and guides the behavior of lower-layer controllers, allowing them to operate autonomously within policy-compliant and \gls{kpi}-safe boundaries. This separation limits the impact of prediction errors and reduces the risk of cascading failures caused by incorrect decisions.

\subsubsection{Monitoring Phase} 
The orchestrator continuously observes energy-related metrics and \gls{kpi} indicators from both the serving cell and neighboring cells. Guardrails are enforced proactively to prevent violations of constraints. Since reacting after a \gls{kpi} violation has occurred is often too late, the orchestrator relies on predicted load and conservative confidence margins to act in advance.
If the monitored indicators suggest elevated risk, such as rising neighbor load, the orchestrator may relax energy-saving configurations, temporarily deactivate certain features, or roll back to a safer operating point. 
}
\begin{figure}
    \centering    \includegraphics[width=1\linewidth]{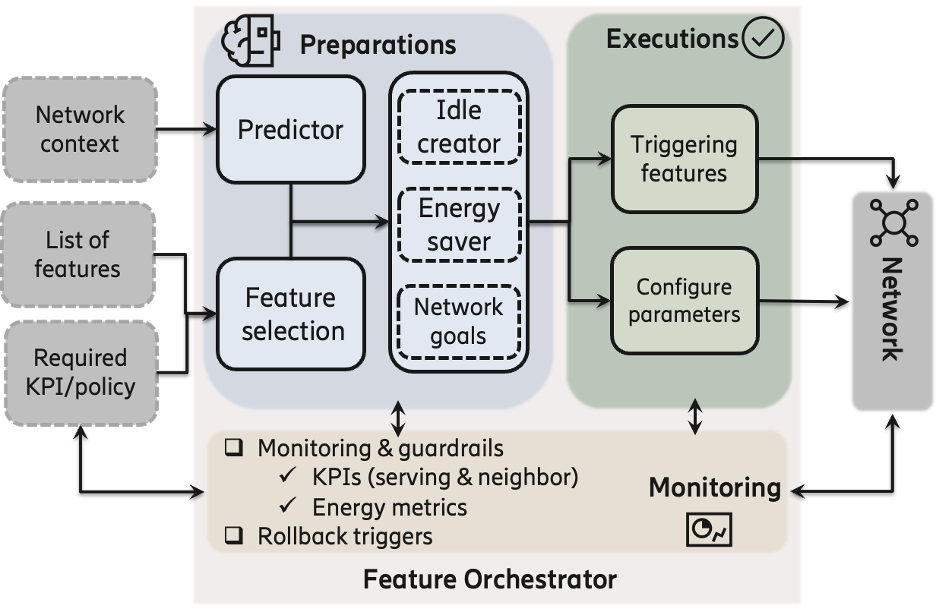}
    \caption{System-level view of the feature orchestrator's  phases for coordinated energy-saving operation.}
    \label{fig:orch}
\end{figure}

{\color{black} Fig.~\ref{fig:orch} illustrates the overall system view of the policy-driven feature orchestrator, including its preparation, execution, and monitoring phases. } Table~\ref{tab:taxonomy} serves as the deployment map. It links each feature to the hardware capabilities it uses, indicates prerequisite coordination with peer features (e.g., steering before shutdown), and annotates the dominant timescale and protocol layer. Practically, it is a checklist for the orchestrator: which enablers must be present, which interfaces must be exposed, and on what timescale the policy should act to achieve \gls{kpi}-safe cost reduction with a focus on energy.

\begin{table*}[t]
\small
\centering
\caption{Energy–performance taxonomy:  {capabilities} vs.\  {features}. Timescales are indicative.}
\label{tab:taxonomy}
\renewcommand\arraystretch{1.15}
\setlength{\tabcolsep}{4pt}
\begin{tabularx}{0.9\textwidth}{l l l l l}
\toprule
\textbf{Item} & \textbf{Category} & \textbf{Layer} & \textbf{Used in} & \textbf{Timescale} \\
\midrule
\multicolumn{5}{l}{\textbf{A. Hardware capabilities (enablers)}}\\
\midrule
Radio sleep capabilities (Micro/Light/Deep) & Capability & \gls{RF}/\gls{phy} & B.3, C.1, C.3 & $\mu$s--tens of ms ($\le$100\,ms)\textsuperscript{1} \\
Multi-band PA hardware                      & Capability & \gls{RF} \gls{FE}  & B.3, C.4            & tens of $\mu$s-ms\textsuperscript{2} \\
DVFS / DRS (compute)                        & Capability & BB/Edge& B.2, C.3, C.5 & $\mu$s--ms\textsuperscript{1} \\
Passive / smart cooling                     & Capability & Site   & standalone     & s--min\textsuperscript{2} \\
\midrule
\multicolumn{5}{l}{\textbf{B. Features that  {create} idle windows}}\\
\midrule
1.\;Lean \gls{nr} design                          & Feature    & \gls{phy}    & B.3, C.1, C.2\textsuperscript{4}       & tens of ms ($\le$160\,ms)\textsuperscript{2} \\
2.\;Architectural split                     & Feature    & Arch   & B.3, B.4, C.2, C.5  & s--min\textsuperscript{3} \\
3.\;Scheduling (control{\,+\,}data)         & Feature    & MAC/\gls{RRC}& C.1, C.3 , C.4          & $\mu$s--100\,ms \textsuperscript{2}\\
4.\;Traffic steering / load placement       & Feature    & RRM    & C.1--C.5       & s--min\textsuperscript{3} \\
5.\;Dynamic spectrum sharing                & Feature    & \gls{phy}/RRM& C.1, C.2, C.4       & ms--tens of ms\textsuperscript{2} \\
\midrule
\multicolumn{5}{l}{\textbf{C. Features that  {utilize} idle windows}}\\
\midrule
1.\;\glspl{asm} (Micro/Light/Deep sleep)    & Feature    & \gls{RF}/RRM & --             & ms--tens of ms\textsuperscript{1} \\
2.\;Carrier shutdown                        & Feature    & \gls{RF}/RRM & --             & ms--s\textsuperscript{1}, s--min\textsuperscript{3} \\
3.\;Massive-\gls{mimo} sleep                      & Feature    & \gls{RF}/MAC & --             & ms--s\textsuperscript{1}, s--min\textsuperscript{3} \\
4.\;Dynamic power pooling                   & Feature    & \gls{RF} \gls{FE}  & --             &  tens of $\mu$s-ms\textsuperscript{2} \\
5.\;CU/DU compute pooling                   & Feature    & Arch   & --             & s--min\textsuperscript{3} \\
\bottomrule
\end{tabularx}

\vspace{3pt}
\begin{minipage}{\textwidth}
\footnotesize
\textsuperscript{1}\,\textbf{Transition time} (required time to ramp down and ramp up the hardware). \quad
\textsuperscript{2}\,\textbf{Control/periodicity} (scheduling or control interval). \quad
\textsuperscript{3}\,\textbf{Control/orchestration} (cluster/cell/carrier orchestration interval).\\
\textsuperscript{4}\,This is due to \gls{ssb}/\gls{sib}-on-demand that can enable more aggressive cell/carrier dormancy. 
 {Numerology note}: slot length scales with subcarrier spacing (e.g., 15\,kHz, 1\,slot is 1\,ms; 30\,kHz, 1\,slot $\approx$ 0.5\,ms), so slot-aligned figures translate accordingly.
\end{minipage}
\end{table*}



\section{Use Case: Coordinated Scheduling and \glspl{asm}}



\begin{figure*}[!t]
\centering
    \begin{subfigure}[t]{0.49\textwidth}
        \centering
        \vspace{0pt}
        {\footnotesize\begin{tikzpicture}
\begin{axis}[%
width=72.0*1.1mm,
height=46.0*1.05mm,
at={(0.769in,0.482in)},
scale only axis,
unbounded coords=jump,
xmin=-0.2,
xmax=8,
xtick={1.15,3.75,6.35},
ytick={30,40,50,60,70,80,90},
xticklabels={{Medium traffic},{Light traffic},{Low traffic}},
ymin=27,
ymax=90,
xlabel style={font=\color{white!15!black}},
ylabel style={font=\color{white!15!black}},
ylabel={Power consumption (relative-power-second)},
axis background/.style={fill=white},
boxplot/draw direction=y,
boxplot/box extend=0.22,
area legend,
ymajorgrids,
boxplot/every average/.style={mark=o, mark size=0.5pt, draw=black, fill=white},
legend plot pos=left,
legend columns=4,
legend style={
    at={(0.5,-0.26)},
    anchor=south,
    legend cell align=left,
    align=left,
    draw=none,
    fill=none,
    font=\scriptsize,
    row sep=1pt,
    /tikz/every even column/.append style={column sep=3pt}
}
]

\pgfplotsset{
  boxplot/every box/.append style={line width=0.2pt},
  boxplot/every median/.append style={line width=0.2pt},
  boxplot/every whisker/.append style={line width=0.3pt},
  base1/.style={fill=baseline_short_col,draw=black},
  base2/.style={fill=baseline_col,draw=black},
  base3/.style={fill=less10_col,draw=black},
  base4/.style={fill=less10_2_col,draw=black},
  base5/.style={fill=less30_col,draw=black},
  base6/.style={fill=less30_2_col,draw=black},
  base7/.style={fill=less60_col,draw=black},
  base8/.style={fill=less60_2_col,draw=black},
}

\addplot[base1,
boxplot prepared={
  lower whisker=78.0373,
  lower quartile=79.6364,
  median=80.4374,
  upper quartile=81.8616,
  upper whisker=87.8851,
  average=80.9126,
}, boxplot/draw position=0.1] coordinates {};
\addlegendentry{\basem}

\addplot[base2,
boxplot/every average/.append style={mark options={fill=white, draw=black}},
boxplot prepared={
  lower whisker=77.2381,
  lower quartile=78.9078,
  median=79.7359,
  upper quartile=81.2288,
  upper whisker=87.6417,
  average=80.2290,
}, boxplot/draw position=0.4] coordinates {};
\addlegendentry{\leanm}

\addplot[base3,
boxplot prepared={
  lower whisker=66.9920,
  lower quartile=69.5866,
  median=70.9020,
  upper quartile=73.2706,
  upper whisker=82.6327,
  average=71.6984,
}, boxplot/draw position=0.7] coordinates {};
\addlegendentry{\tgm{10}}

\addplot[base4,
boxplot prepared={
  lower whisker=66.9158,
  lower quartile=69.5104,
  median=70.8258,
  upper quartile=73.1945,
  upper whisker=82.5565,
  average=71.6222,
}, boxplot/draw position=1.0] coordinates {};
\addlegendentry{\tgd{10}}

\addplot[base5,
boxplot prepared={
  lower whisker=66.9632,
  lower quartile=69.5780,
  median=70.8346,
  upper quartile=72.6865,
  upper whisker=77.7667,
  average=71.2304,
}, boxplot/draw position=1.3] coordinates {};
\addlegendentry{\tgm{30}}

\addplot[base6,
boxplot prepared={
  lower whisker=66.8870,
  lower quartile=69.5053,
  median=70.7598,
  upper quartile=72.6152,
  upper whisker=77.6906,
  average=71.1577,
}, boxplot/draw position=1.6] coordinates {};
\addlegendentry{\tgd{30}}

\addplot[base7,
boxplot prepared={
  lower whisker=66.5945,
  lower quartile=67.5433,
  median=67.8492,
  upper quartile=68.0754,
  upper whisker=68.6591,
  average=67.7789,
}, boxplot/draw position=1.9] coordinates {};
\addlegendentry{\tgm{60}}

\addplot[base8,
boxplot prepared={
  lower whisker=42.3379,
  lower quartile=45.2836,
  median=46.0188,
  upper quartile=46.4721,
  upper whisker=47.2052,
  average=45.7663,
}, boxplot/draw position=2.2] coordinates {};
\addlegendentry{\tgd{60}}

\addplot[base1, forget plot,
boxplot prepared={
  lower whisker=74.3800,
  lower quartile=75.2794,
  median=75.7455,
  upper quartile=76.5721,
  upper whisker=80.0274,
  average=76.0236,
}, boxplot/draw position=2.7] coordinates {};

\addplot[base2, forget plot,
boxplot prepared={
  lower whisker=73.2598,
  lower quartile=74.2289,
  median=74.7271,
  upper quartile=75.5879,
  upper whisker=79.2573,
  average=75.0027,
}, boxplot/draw position=3.0] coordinates {};

\addplot[base3, forget plot,
boxplot prepared={
  lower whisker=62.3384,
  lower quartile=63.8945,
  median=64.6080,
  upper quartile=66.0590,
  upper whisker=71.6342,
  average=65.0949,
}, boxplot/draw position=3.3] coordinates {};

\addplot[base4, forget plot,
boxplot prepared={
  lower whisker=62.2623,
  lower quartile=63.8183,
  median=64.5281,
  upper quartile=65.9829,
  upper whisker=71.5581,
  average=65.0165,
}, boxplot/draw position=3.6] coordinates {};

\addplot[base5, forget plot,
boxplot prepared={
  lower whisker=62.3358,
  lower quartile=63.8762,
  median=64.5834,
  upper quartile=66.0105,
  upper whisker=71.3014,
  average=65.0815,
}, boxplot/draw position=3.9] coordinates {};

\addplot[base6, forget plot,
boxplot prepared={
  lower whisker=62.2416,
  lower quartile=63.7610,
  median=64.4752,
  upper quartile=65.9130,
  upper whisker=71.2253,
  average=64.9664,
}, boxplot/draw position=4.2] coordinates {};

\addplot[base7, forget plot,
boxplot prepared={
  lower whisker=62.2455,
  lower quartile=63.8101,
  median=64.5480,
  upper quartile=65.7133,
  upper whisker=67.6846,
  average=64.7553,
}, boxplot/draw position=4.5] coordinates {};

\addplot[base8, forget plot,
boxplot prepared={
  lower whisker=33.8709,
  lower quartile=35.8728,
  median=37.2092,
  upper quartile=39.9286,
  upper whisker=45.4725,
  average=38.0262,
}, boxplot/draw position=4.8] coordinates {};

\addplot[base1, forget plot,
boxplot prepared={
  lower whisker=66.1872,
  lower quartile=66.5910,
  median=66.7631,
  upper quartile=67.0255,
  upper whisker=68.2795,
  average=66.8320,
}, boxplot/draw position=5.3] coordinates {};

\addplot[base2, forget plot,
boxplot prepared={
  lower whisker=64.8352,
  lower quartile=65.2308,
  median=65.3975,
  upper quartile=65.6500,
  upper whisker=66.9147,
  average=65.4714,
}, boxplot/draw position=5.6] coordinates {};

\addplot[base3, forget plot,
boxplot prepared={
  lower whisker=57.3914,
  lower quartile=57.8032,
  median=58.0149,
  upper quartile=58.3649,
  upper whisker=59.8791,
  average=58.1233,
}, boxplot/draw position=5.9] coordinates {};

\addplot[base4, forget plot,
boxplot prepared={
  lower whisker=55.7777,
  lower quartile=56.3773,
  median=56.5941,
  upper quartile=56.9316,
  upper whisker=58.5594,
  average=56.7000,
}, boxplot/draw position=6.2] coordinates {};

\addplot[base5, forget plot,
boxplot prepared={
  lower whisker=57.2378,
  lower quartile=57.6753,
  median=57.8944,
  upper quartile=58.2733,
  upper whisker=60.4669,
  average=58.0543,
}, boxplot/draw position=6.5] coordinates {};

\addplot[base6, forget plot,
boxplot prepared={
  lower whisker=54.9167,
  lower quartile=55.5340,
  median=55.7557,
  upper quartile=56.1584,
  upper whisker=59.4135,
  average=55.9750,
}, boxplot/draw position=6.8] coordinates {};

\addplot[base7, forget plot,
boxplot prepared={
  lower whisker=57.2301,
  lower quartile=57.6355,
  median=57.8447,
  upper quartile=58.2102,
  upper whisker=59.8008,
  average=57.9858,
}, boxplot/draw position=7.1] coordinates {};

\addplot[base8, forget plot,
boxplot prepared={
  lower whisker=27.8048,
  lower quartile=28.1990,
  median=28.3494,
  upper quartile=28.7119,
  upper whisker=31.6306,
  average=28.5625,
}, boxplot/draw position=7.4] coordinates {};

\end{axis}
\end{tikzpicture}}
        \caption{}
        \label{fig:energyCons}
    \end{subfigure}
    \hfill
    \begin{subfigure}[t]{0.49\textwidth}
        \centering
        \vspace{0pt}
        {\footnotesize\begin{tikzpicture}
\begin{axis}[%
width=72.0*1.1mm,
height=46.0*1.05mm,
at={(0.769in,0.482in)},
scale only axis,
unbounded coords=jump,
xmin=0.1,
xmax=5.3,
xtick={1,2.7,4.4},
xticklabels={{Medium traffic},{Light traffic},{Low traffic}},
ytick={0,0.2,0.4,0.6,0.8,1,1.2,1.4,1.6},
ymin=0,
ymax=1.6,
xlabel style={font=\color{white!15!black}},
ylabel style={font=\color{white!15!black}},
ylabel={DL PDCP Throughput (Mb/s)},
axis background/.style={fill=white},
boxplot/draw direction=y,
boxplot/box extend=0.22,
area legend,
ymajorgrids,
boxplot/every average/.style={mark=o, mark size=0.5pt, draw=black, fill=white},
legend plot pos=left,
legend columns=3,
legend style={
    at={(0.5,-0.09)},
    anchor=north,
    legend cell align=left,
    align=left,
    draw=none,
    fill=none,
    font=\scriptsize,
    row sep=1pt,
    /tikz/every even column/.append style={column sep=4pt}
}
]
\pgfplotsset{
  boxplot/every box/.append style={line width=0.2pt},
  boxplot/every median/.append style={line width=0.2pt},
  boxplot/every whisker/.append style={line width=0.3pt},
  base1/.style={fill=baseline_short_col,draw=black},
  base2/.style={fill=baseline_col,draw=black},
  base3/.style={fill=less10_col,draw=black},
  base4/.style={fill=less30_col,draw=black},
  base5/.style={fill=less60_col,draw=black},
}

\addplot+[base1,
boxplot prepared={
  lower whisker=1.11334,
  lower quartile=1.27098,
  median=1.30216,
  upper quartile=1.33359,
  upper whisker=1.48965,
  average=1.30262,
}, boxplot/draw position=0.4] coordinates {};
\addlegendentry{\base}

\addplot+[base2,
boxplot prepared={
  lower whisker=1.11356,
  lower quartile=1.27130,
  median=1.30246,
  upper quartile=1.33385,
  upper whisker=1.48981,
  average=1.30288,
}, boxplot/draw position=0.7] coordinates {};
\addlegendentry{\lean}

\addplot+[base3,
boxplot prepared={
  lower whisker=1.11356,
  lower quartile=1.27116,
  median=1.30243,
  upper quartile=1.33379,
  upper whisker=1.48990,
  average=1.30276,
}, boxplot/draw position=1.0] coordinates {};
\addlegendentry{\tg{10}}

\addplot+[base4,
boxplot prepared={
  lower whisker=0.452529,
  lower quartile=1.26969,
  median=1.30158,
  upper quartile=1.33323,
  upper whisker=1.48746,
  average=1.29835,
}, boxplot/draw position=1.3] coordinates {};
\addlegendentry{\tg{30}}

\addplot+[base5,
boxplot prepared={
  lower whisker=0.126489,
  lower quartile=1.04913,
  median=1.24488,
  upper quartile=1.29843,
  upper whisker=1.45178,
  average=1.14290,
}, boxplot/draw position=1.6] coordinates {};
\addlegendentry{\tg{60}}

\addplot[base1, forget plot,
boxplot prepared={
  lower whisker=0.629604,
  lower quartile=0.746098,
  median=0.769782,
  upper quartile=0.793632,
  upper whisker=0.916066,
  average=0.770126,
}, boxplot/draw position=2.1] coordinates {};

\addplot[base2, forget plot,
boxplot prepared={
  lower whisker=0.628998,
  lower quartile=0.746808,
  median=0.770518,
  upper quartile=0.794291,
  upper whisker=0.916906,
  average=0.770803,
}, boxplot/draw position=2.4] coordinates {};

\addplot[base3, forget plot,
boxplot prepared={
  lower whisker=0.628616,
  lower quartile=0.746504,
  median=0.770181,
  upper quartile=0.794007,
  upper whisker=0.916330,
  average=0.770551,
}, boxplot/draw position=2.7] coordinates {};

\addplot[base4, forget plot,
boxplot prepared={
  lower whisker=0.628924,
  lower quartile=0.746452,
  median=0.770128,
  upper quartile=0.794035,
  upper whisker=0.916621,
  average=0.770503,
}, boxplot/draw position=3.0] coordinates {};

\addplot[base5, forget plot,
boxplot prepared={
  lower whisker=0.222182,
  lower quartile=0.744587,
  median=0.769002,
  upper quartile=0.793002,
  upper whisker=0.916715,
  average=0.766654,
}, boxplot/draw position=3.3] coordinates {};

\addplot[base1, forget plot,
boxplot prepared={
  lower whisker=0.129191,
  lower quartile=0.188938,
  median=0.201120,
  upper quartile=0.213641,
  upper whisker=0.269178,
  average=0.201513,
}, boxplot/draw position=3.8] coordinates {};

\addplot[base2, forget plot,
boxplot prepared={
  lower whisker=0.129356,
  lower quartile=0.189533,
  median=0.201742,
  upper quartile=0.214349,
  upper whisker=0.269810,
  average=0.202158,
}, boxplot/draw position=4.1] coordinates {};

\addplot[base3, forget plot,
boxplot prepared={
  lower whisker=0.129301,
  lower quartile=0.189191,
  median=0.201400,
  upper quartile=0.213940,
  upper whisker=0.269870,
  average=0.201787,
}, boxplot/draw position=4.4] coordinates {};

\addplot[base4, forget plot,
boxplot prepared={
  lower whisker=0.0817891,
  lower quartile=0.189247,
  median=0.201513,
  upper quartile=0.214018,
  upper whisker=0.269784,
  average=0.201821,
}, boxplot/draw position=4.7] coordinates {};

\addplot[base5, forget plot,
boxplot prepared={
  lower whisker=0.105551,
  lower quartile=0.189053,
  median=0.201272,
  upper quartile=0.213809,
  upper whisker=0.269360,
  average=0.201651,
}, boxplot/draw position=5.0] coordinates {};

\end{axis}
\end{tikzpicture}%
}
        \caption{}
        \label{fig:thrBox}
    \end{subfigure}
    \caption{%
    {\color{black}The impact of sleep modes, scheduling, and \gls{nr} lean design under different traffic scenarios on the empirical distributions of (a) power consumption and (b) \gls{dl} SDU PDCP throughput. Each box spans the 25th--75th percentiles, the horizontal line denotes the median, the circle denotes the mean, and the whiskers indicate the minimum and maximum values.}}
    \label{fig:kpi}
    \vspace{-3mm}
\end{figure*}
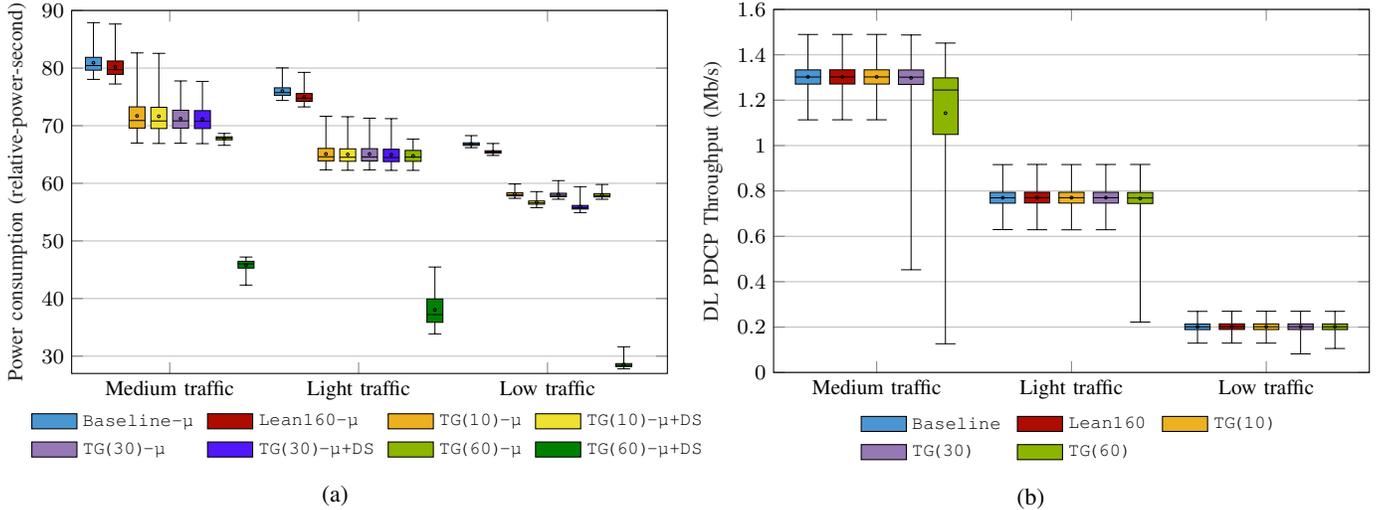


\begin{table}[htb] \small
\centering
\caption{Power states of \gls{gnb} and simulation parameters}
\label{tab:powerModel_sim}
\setlength{\tabcolsep}{4pt}
\renewcommand{\arraystretch}{1.1}
\begin{tabularx}{\columnwidth}{l*{6}{>{\centering\arraybackslash}X}}
\toprule
\multicolumn{7}{c}{\textbf{Simulation parameters}} \\
\midrule
\multicolumn{2}{l}{Bandwidth} & \multicolumn{1}{c}{$20$\,MHz}
& \multicolumn{2}{l}{Frequency} & \multicolumn{1}{c}{$3.5$\,GHz} \\
\multicolumn{2}{l}{Subcarrier spacing} & \multicolumn{1}{c}{$30$\,kHz}
& \multicolumn{2}{l}{Slot duration} & \multicolumn{1}{c}{$0.5$\,ms} \\
\multicolumn{2}{l}{Power} & \multicolumn{1}{c}{$40$\,W}
& \multicolumn{2}{l}{Num. UEs} & \multicolumn{1}{c}{20} \\
\multicolumn{2}{l}{TX/RX chains} & \multicolumn{1}{c}{16}
& \multicolumn{2}{l}{Num. samples} & \multicolumn{1}{c}{$10^6$} \\
\midrule
 & \multicolumn{2}{c}{\textbf{OFF States}} & \multicolumn{4}{c}{\textbf{ON States}}\\
\cmidrule(r){2-3} \cmidrule(r){4-7}
\textbf{Power state} & Deep sleep & Micro sleep & Active TX & Idle TX & Active RX & Idle RX \\
\cmidrule(r){2-3} \cmidrule(r){4-7}
\textbf{Relative power} & 1 & 55 & 119.3 & 71.3 & 80.33 & 70.2 \\
\bottomrule
\end{tabularx}
\end{table}

We evaluate the energy–throughput trade-off of coordinated idle-time creation using traffic gating, \gls{nr} lean signaling, and \glspl{asm} on top of the radio sleep capability in Section~\ref{sec:hw}.
We use an event-driven, symbol-level 6G/\gls{nr} simulator aligned with 3GPP assumptions; key parameters and the normalized power-state model ~\cite{3GPPTraffic} are given in Table~\ref{tab:powerModel_sim}. All power levels were normalized to the deep‑sleep state. 
Only \gls{dl} traffic was generated, where
traffic demand followed the 3GPP profiles in~\cite{3GPPTraffic}. Three traffic profiles were defined in terms of mean \gls{prb} utilization:
\begin{itemize}
    \item \textbf{Low traffic}: $6.5\%$ (matches $0 {<} \mathrm{load}{\leq} 15$\% in~\cite{3GPPTraffic})
    \item \textbf{Light traffic}: $25\%$ (matches $15\% {<}\mathrm{load}{\leq} 30$\% in~\cite{3GPPTraffic})
    \item \textbf{Medium traffic}: $42\%$ (matches $30\% {<}\mathrm{load}{\leq} 50$\% in~\cite{3GPPTraffic})
\end{itemize}
Note that \gls{prb} utilization fluctuated over time, and the values above refer to the respective time-averaged levels.
Following~\cite{3GPPTraffic}, deep sleep incurred a $50$\,ms transition time for ramping up and ramping down the corresponding hardware, each costing $1$ relative‑power‑second. The transition time and energy for micro sleep were assumed to be negligible. {\color{black} While the evaluation focuses on a single carrier to isolate coordination effects, multi-carrier multi-cell deployments with mobility provide additional degrees of freedom, such as steering latency-sensitive traffic to always-on carriers while aggressively gating others. This may mitigate latency–throughput trade-offs but requires tighter orchestration across carriers.} {\color{black} Moreover, saving energy can affect neighboring cell \glspl{kpi} such as load. Therefore, the orchestrator should consider neighbor-aware guardrails for feature orchestration.}


For our case study, we consider the following benchmarks:
\begin{itemize}
    \item \base: Vanilla simulations using the default signaling periodicities: \gls{prach} $5$\,ms, \gls{ssb} $20$\,ms, \gls{csirs} $5$\,ms, \gls{bcch}/SIB1 $50$\,ms.
    \item \lean: Same as \base, but every signaling periodicity above is raised to $160$\,ms, realizing the \gls{nr} lean design principle to cut always-on transmissions.
    \item \tg{$\tau$}: Builds on \lean\! and adds traffic-gating: the scheduler withholds non-priority traffic for a fixed interval $\tau {\in} \{10, 30, 60\}$\,ms, releasing it when the timer expires or a high-priority traffic arrives. This extension prolongs consecutive idle times and improves the opportunity for deeper energy-saving modes.
\end{itemize}
Each benchmark may appear in two variants: \texttt{-\textmu}, which assumes the \gls{ru} supports only micro sleep, and \dsdash, which assumes the \gls{ru} can enter \textbf{both} micro and deep sleep (e.g., \leand). {\color{black} The {\ds}  provides a lower bound on energy consumption enabled by coordination, assuming perfect predictions and optimal decisions with no \gls{kpi} degradation (due to deep sleep).
In practice, a short-horizon prediction provides such information.} 

{\color{black} \figurename\,\ref{fig:energyCons} demonstrates the distribution of power consumption, summarizing $500$ independent simulation seeds per benchmark. The \ds\, distributions represent a lower bound: using offline knowledge of whether the preceding ${\geq}50$\,ms interval was idle, we account the \gls{gnb} as being in deep sleep. Thus, the results assume perfect short-horizon idleness prediction and optimal timing. However, with prediction errors, the achievable savings may reduce and/or \glspl{kpi} may degrade.
}
In both \base\, and \lean, micro- and deep-sleep benchmarks yield identical power consumption because neither benchmark attains the ${\geq}50$\,ms period required for deep sleep.
\figurename\,\ref{fig:energyCons} delivers four main takeaways:
\begin{itemize}
    \item  \textbf{Solely extending signaling periods yields little gains}: With all periodicities stretched to $160$\,ms (\lean), mean power drops by up to $2$\% because of the micro sleeps (under the Low traffic profile) versus \base. Such an extension provides a sufficient interval to activate deep sleep. However, traffic arrival becomes the limiting factor. Hence, cutting signaling overhead is necessary but not sufficient for meaningful savings.
    
    \item \textbf{Traffic gating helps even with micro sleep alone}: A $10$\,ms gating window (\tgm{10}) cuts mean power by roughly 10\% at medium traffic. Expanding the window to 60\,ms brings further benefit at medium traffic but little extra at Low or Light loads, where idle gaps are already long enough.
    
    \item \textbf{Deep sleep is the real lever}: Allowing deep sleep in \tgd{60} slashes power consumption across all traffic regimes, easily outperforming every micro-sleep-only strategy.

    \item \textbf{Savings depend on coordinated features}: Significant gains appear only when extended signaling, traffic gating, and deep sleep work cooperatively. Assuming deep sleep, \lean\,with no traffic illustrates the most energy savings. However, under Low traffic, gating becomes important to manage the traffic for energy savings (e.g., a $30$\,ms gate offers virtually no advantage over $10$\,ms). 
\end{itemize}

\figurename\,\ref{fig:thrBox} shows the distribution of \gls{pdcp} \gls{sdu} throughput for all benchmarks, and thus reveals the performance cost of the power savings seen in \figurename\,\ref{fig:energyCons}. Under the Low traffic profile, \tg{60} leaves every throughput statistic unchanged, even though it delivers the largest power reduction (i.e., $58$\% compared to \base). In the Light traffic profile, the same gating timer again preserves the mean and all upper percentiles but decreases the minimum throughput from $0.66$\,Mb/s (\base) to $0.26$\,Mb/s. By contrast, the Medium traffic profile shows a noticeable reduction of the entire distribution, confirming that long gating windows can result in reduced throughput at higher loads. These results show that substantial power savings can be realized with negligible throughput impact when the cell operates in Light or Low traffic, and the trade-off becomes significant only as traffic approaches the Medium level. Gating may cause extra latency, however, a large fraction of mobile traffic is latency-tolerant (e.g., buffered video streaming), for which \tg{60} is acceptable. 
Moreover, since there are typically multiple configured carriers on a \gls{gnb}, latency-sensitive flows can be steered to other carriers, allowing some carriers safely apply longer gating windows.


\vspace{-10pt}
\section{Open Issues}
We summarize four open issues arising from feature design and orchestration.
\vspace{-10pt}
\subsection{Energy Consumption Observability}
Achieving energy efficiency relies on predicting and orchestrating when and where idle windows exist, and estimating their payoff. Nevertheless, the demands of a single \gls{rat} or service may block deeper sleep modes activation, and differing traffic shares across digital, transport, and radio layers further complicate energy management among \glspl{rat}/services. Today, energy consumption is typically measured using meters installed on physical nodes, whose processing resources are shared by multiple logical nodes (e.g., co-located cells or \glspl{gnb}).
However, feature optimizations often require finer-grained energy measurements (e.g., carrier-level measurements).
This misalignment complicates attribution (as node-level measurements often fail to map to cell-level energy) and can obscure energy-performance trade-offs (e.g., Fig.~\ref{fig:kpi}). Promising directions include finer-granularity energy counters and a consistent split between static and load-dependent power. 


\subsection{Traffic Dynamics}
Network traffic, service requirements, and the network status determine the required resources for operation. Network traffic can be modeled as the \gls{ue}-aggregated data served in discrete sessions, each containing one or multiple \gls{ue} payloads. We can categorize the sessions into three classes: 
i) large (${>}20$\,MB forming $\approx 2\%$ of total sessions) that demand intensive radio resources (e.g., high number of \glspl{prb}, transmit power),
ii) medium ($1$–$20$\,MB forming $\approx 3\%$ of sessions) that require moderate resources, and 
iii) small (${<}1$\,MB forming $\approx 95\%$ of traffic) that need minimal radio resources. Networks are designed around large sessions even though most sessions are small, driving disproportionate energy consumption to provision a minority traffic class. Additionally, short sessions may puncture otherwise viable sleep windows. This contrasts with the industry's aim for zero energy at zero load.  To address this issue, scheduling should become session-class aware and plan over short-horizon  to create and preserve idle windows. Further study is needed to manage this traffic mix more energy-efficiently.


\vspace{-10pt}
\subsection{Conflict Management}
Energy-saving features are typically designed locally (per node or function), yet dependencies on other network functions can create unintended interactions. For example, interactions among \gls{dss}, \glspl{asm}, carrier shutdown, and massive \gls{mimo} sleep can negate gains or degrade \glspl{kpi} when activated together.
Conflicts often arise from overlapping timers, incompatible prerequisites, or “ping-pong” between policies. Conflict management can be realized locally or via a higher-level orchestrator. Such control mechanisms may include pre-activation “what-if” evaluation to further reduce risk, devising a policy graph, or post-activation rollbacks.


\vspace{-10pt}
\subsection{\Gls{ml} for Energy Efficiency}
{\color{black}

\subsubsection{Data availability and observability}
Training models for energy-aware decisions requires accurate and timely measurements of energy consumption. However, energy is often measured at coarse aggregation points, such as the site or \gls{ru} level while energy-saving actions are triggered at finer granularity, such as carrier, or cell. Where to place energy probes, how to aggregate measurements across layers, and what temporal granularity is sufficient for learning remain open questions. If measurements are delayed or too coarse, the resulting models become practically ineffective.

\subsubsection{Computational resources and latency} Energy-saving features inherently trade latency for energy-saving. \Gls{ai}-based orchestration adds additional delay due to data pre-processing, inference, and control signaling. Even if inference itself is fast, the cumulative latency may reduce the usable idle window or increase the risk of \gls{kpi} degradation. Therefore, models must be designed with minimal computational footprint, placed close to the decision point, and explicitly account for computation-related latency when triggering actions.

\subsubsection{Delayed  impact of  actions}
Unlike throughput or scheduling decisions, the impact of an energy-saving action is often observed only after some time, when energy consumption is measured over an interval. Moreover, observed changes in energy or \glspl{kpi} may be influenced by reasons such as traffic variations, and mobility. This makes it difficult to attribute outcomes to specific actions, complicating model training, monitoring, and explainability. This requires model training and evaluation to consider delayed effects, uncertainty, and partial observability. \Gls{ml} techniques such as offline evaluation, and horizon-aware solutions (e.g., conservative \gls{rl}/bandits) are among the approaches addressing this issue.


}

\section{Conclusion}

Unlocking the full energy-saving potential of the \gls{ran} requires not only capable hardware but also tight coordination among interconnected features. In this article, we first provided a clear taxonomy of capabilities and features, separating those that create idle windows from those that utilize them, and then discussed how a single logical coordination node enables cooperation to improve energy saving.
Our end-to-end results, conducted using a 3GPP-aligned simulator, confirmed that combining extended lean-\gls{nr} signaling, traffic gating, and predictive wakeups with deep sleep delivers substantial energy savings while keeping throughput at a reasonable level. The use case study further indicates that well-orchestrated features working in harmony are a necessity for end-to-end energy-efficient cost-effective networks.






\ifCLASSOPTIONcaptionsoff
  \newpage
\fi

\bibliographystyle{IEEEtran}
\bibliography{components/bib/IEEEabrv, components/ref.bib}

\end{document}